\documentclass[onecolumn,aps,superscriptaddress,nofootinbib]{revtex4}
\usepackage{graphicx,amsmath}

\usepackage{graphicx}
\usepackage{dcolumn}
\usepackage{bm}

\makeatletter
\renewcommand \thesection {\@arabic\c@section}
\makeatother
\makeatletter 
\@addtoreset{equation}{section}
\makeatother  

\begin{document}


\title{Some Comments on the Holographic Heavy Quark Potential in a Thermal Bath}

\author{Yan Wu}
 \email{wuyan@iopp. ccnu. edu. cn}
\affiliation {Institute of Particle Physics,Key Laboratory of QLP
of  MOE, Huazhong Normal University, Wuhan 430079, China}
\author{ De-fu  Hou}%
 \email{hdf@iopp. ccnu. edu. cn}
 \affiliation{Institute of Particle Physics,Key Laboratory of QLP of  MOE, Huazhong Normal
University, Wuhan 430079, China}
\author{Hai-cang Ren}
\email{ren@mail.rockefeller.edu}   \affiliation{Physics
Department, The Rockefeller University, 1230 York Avenue,  New
York,  NY 10021-6399} \affiliation{Institute of Particle
Physics,Key Laboratory of QLP of  MOE, Huazhong Normal University,
Wuhan 430079, China}

\begin{abstract}
The heavy quark potential of a thermal Yang-Mills theory in strong
coupling limit is explored in terms of the holographic principle.
With a fairly general AdS/QCD metric the heavy quark potential
displays a kink-like screening in the plasma phase. This behavior
may conflict the causality of a field theory that is
mathematically equivalent to the thermal Yang-Mills.
\end{abstract}
\maketitle
\section{Introduction}
The heavy-quark potential (the interaction energy between a quark
and its antiparticle in the infinite mass limit) is a very
important quantity of QCD. Not only does it provide the
information of the quarkonium dissociation which signals the
formation of QGP in heavy ion collisions\cite{Satz}, but also is
one of the basic probes to explore the phase diagram with nonzero
temperature and baryon density. While the potential is Coulomb
like for a short separation, $r<<\Lambda_\text{QCD}^{-1}$, the
potential rises linearly at large distance,
$r>>\Lambda_\text{QCD}^{-1}$ \cite{Eichten}in the confined phase
without light quarks as is suggested by the Regge behavior of
meson spectra and is expected to be screened within a radius of
the order $T^{-1}$ in the deconfined phase.

Mathematically, the heavy-quark potential can be extracted from
the expectation value of a Wilson loop, or the correlator between
two Polyakov loops.\cite{Eichten1} The highly nonperturbative
nature of infrared QCD makes it analytically intractable,
especially in the confined phase. The lattice simulation of the
Wilson loops and/or the Polyakov loops has accumulated sufficient
evidence supporting the linear potential below the deconfiment
temperature and the screened potential in the plasma
phase\cite{Lattice1, Lattice11, Lattice111, Lattice1111,
Lattice, Lattice11111, Lattice3, HSatz, Hatsuda-lattice}. Furthermore a resummation
of perturbative series yields an exponentially screening at high
temperature in weak coupling \cite{pisarski,weldon}.

The advent of the holographic principle \cite{hooft,susskind},
especially the AdS/CFT correspondence\cite{Maldacena} opens a new
avenue towards analytic treatments of the strong coupling limit of
a gauge theory, in particular, the $N=4$ supersymmetric $SU(N_c)$
Yang-Mills theory at large $N_c$ and large 't Hooft coupling,
$\lambda\equiv N_cg_\text{YM}^2$ with $g_{\text{YM}}$ the
Yang-Mills coupling. The heavy-quark potential\cite{Maldacena,
Brandhuber, Rey, Chu, Foroni,Zhang, Huang1, cai, Huang,Noronha,
Jeon, Jeon1, Hatsuda, Willy} at zero temperature is Coulomb like
with the strength proportional to
$\sqrt{\lambda}$.\cite{Maldacena, Chu, Foroni} At a nonzero
temperature,\cite{Brandhuber, Rey, Zhang} the potential displays a
kink-like screening with a radius of $r_s\simeq 0.754(\pi
T)^{-1}$, i.e. the potential is flattened out beyond $r_s$. This
transition is interpreted as string melting in \cite{Brandhuber}
and will be referred to as the kink-like screening in this paper.
But the super Yang-Mills is not QCD and the conformal property of
the former makes the Coulomb like behavior the only possible
outcome at zero temperature following a dimensional argument. A
cousin of AdS/CFT with an infrared cutoff, AdS/QCD, has been
actively investigated and is able to provide a linear heavy
quark-potential at zero temperature. The nonzero temperature
behavior were found to fit lattice data quite well.\cite{Huang1,
cai, Huang}

In this paper, we shall explore analytically the screening
property of the heavy-quark potential within the framework of
AdS/QCD. In the next section, the heavy-quark potential of the
$N=4$ super Yang-Mills at a nonzero temperature following AdS/CFT
correspondence will be reviewed, where we shall also set up our
notations. In the section III, we shall show that under a fairly
general conditions of the metric underlying AdS/QCD, the screening
remains kink-like, like that of the super Yang-Mills. In other
words, AdS/QCD cannot provide a exponentially screening potential
in the plasma phase. Different scenarios of a smooth screening
behaviors proposed in the literature will be discussed in the
section IV, where we shall also point out a potential gap between
the kink-like screening potential and the fundamental principles
of quantum field theories.

\section{The Heavy-Quark Potential in a $N=4$ Super Yang-Mills Plasma in Strong Coupling}

AdS/CFT correspondence relates the type IIB superstring theory in
$AdS_5\times S^5$ background to the $N=4$ supersymmetric $SU(N_c)$
Yang-Mills theory at the AdS boundary. In particular, the thermal
expectation value of a Wilson loop $C$, $<W[C]>$, at large $N_c$
and large 't Hooft coupling constant $\lambda\equiv N_c
g_\text{YM}^2$ corresponds to the minimum area of a string world
sheet in the Schwarzschild-$AdS_5$ metric,
\begin{eqnarray}
ds^2=\frac{1}{z^2}\bigg[\left(1-\frac{z^4}{z_h^4}\right)d\tau^2+\left(1-\frac{z^4}{z_h^4}\right)^{-1}dz^2+d\vec{x}^2\bigg]
\label{Sch_AdS}
\end{eqnarray}
bounded by the loop $C$ at the boundary $z=0$, i.e.
\begin{equation}
<W[C]> = \text{const.}e^{-\text{min.}(S_\text{NB}[C])}
\label{leading}
\end{equation}
where the Nambu-Goto action
\begin{equation}
S_\text{NB}[C] = \frac{{\cal S}}{2\pi\alpha¡¯} =
\frac{1}{2\pi\alpha¡¯}\int d^2\sigma\sqrt{g}
\end{equation}
with ${\cal S}$ the world sheet area, and $g$ the determinant of
the induced world sheet metric, $ds^2=g_{ab}d\sigma^a d\sigma^b$
($a,b=0,1$). The horizon radius $z_h$ corresponds to the
temperature $T=(\pi z_h)^{-1}$ and the string tension corresponds
to the 't Hooft coupling $\sqrt{\lambda}=\alpha^{-1}$.

The Wilson loop for the free energy of a single heavy quark or
antiquark is a Polyakov line winding up the Euclidean time
dimension and will be denoted by ${\cal C}_1$. The Wilson loop for
the free energy of a pair of heavy quark and antiquark consists of
two Polyakov lines running in opposite directions and will be
denoted by ${\cal C}_2$. The heavy quark potential $V(r)$
corresponds to the interaction part of the free energy of the
quark pair and is extracted from the ratio
\begin{equation}
\frac{<W[{\cal C}_2]>}{|<W[{\cal C}_1]>|^2} =\frac{\text{Tr}<{\cal
P}(\vec r){\cal P}^\dagger(0)>}{|\text{Tr}<{\cal P}(0)>|^2} \equiv
e^{-\frac{V(r)}{T}} \label{def}
\end{equation}
where the Wilson line operator \footnote{Strictly speaking, the
"heavy quark" in the super Yang-Mills refers to the heavy W-boson
of the symmetric breaking from $SU(N_c)$ to $SU(N_c-1)$ and the
Polyakov line operator contains the contribution from the scalar
field. See e.g. \cite{Maldacena} for details.}
\begin{equation}
{\cal P}(\vec r)={\cal T}e^{-i\int_0^{T^{-1}}dtA_0(\vec r,\tau)}
\end{equation}
with $A_0(\vec r,\tau)$ the temporal component of the Lie Algebra
valued gauge potential and ${\cal T}$ the time ordering operator.
The solution of the Euler-Lagrange equation for ${\cal C}_1$ is a
world sheet with constant 3D spatial coordinates $\vec x$
extending from the AdS boundary $z=0$ to the black hole horizon
$z=z_h$ and its NG action will be denoted by $\frac{1}{T}{\cal
A}_1$. The solution of the Euler-Lagrange equation for ${\cal
C}_2$ can be either a connected nontrivial world sheet shown in
Fig.1a or two parallel world sheets shown in Fig.1b with each
identical to the world sheet of ${\cal C}_1$.
\begin{figure}[!htb]
\centering \centering
\includegraphics[scale=0.52, clip=true]{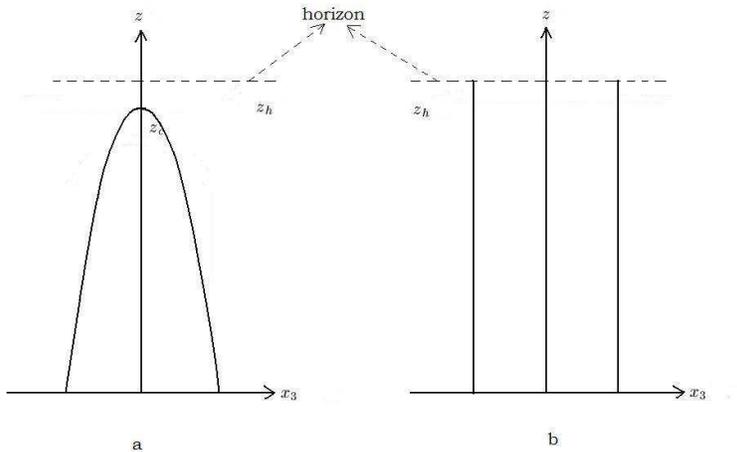}
\hspace{0in} \caption{(a) The connected nontrivial world sheet
from the boundary $z=0$ extending to $z_c<z_h$. (b) is two
parallel world sheets starting from the boundary $z=0$ and ending
at the horizon $z=z_h$. The upper dashed lines represent the black
hole horizon.}
\end{figure}
The NG action of the former will be denoted by $\frac{1}{T}{\cal
A}_2$ while that of the latter is given by $\frac{2}{T}{\cal A}_1$
and the minimum of them contributes to the free energy. It follows
from (\ref{def}) that
\begin{equation}
V(r)=\text{min}({\cal A},0) \label{kink}.
\end{equation}
where ${\cal A}\equiv {\cal A}_2-2{\cal A}_1$ and this combination
cancels the UV divergences pertaining to ${\cal A}_1$ and ${\cal
A}_2$. We shall name ${\cal A}(r)$ as the candidate potential.

For $N=4$ super Yang-Mills, the Euler-Lagrange equation of
minimizing $S_\text{NB}[C]$ yields the following parametric form
of the function candidate potential ${\cal A}(r)$\cite{Maldacena,
Rey, Brandhuber}
\begin{equation}
\left\{\begin{array}{ll}
r=2\sqrt{z_h^4-z_c^4}\int_0^{z_c}dz\frac{z^2}{\sqrt{(z_h^4-z^4)(z_c^4-z^4)}}\\
{\cal A}=\frac{\sqrt{\lambda z_c^2}}{\pi
z_h^2}\bigg[\int_0^{z_c}\frac{dz}{z^2}\left(\sqrt{\frac{z_h^4-z^4}{z_c^4-z^4}}-1\right)
-\int_0^{z_h} dz\frac{1}{z^2}\bigg]\end{array}\right. \label{V}
\end{equation}
where the parameter $z_c$ is the maximum extension of the world
sheet in the bulk. The candidate potential ${\cal A}$ as a
function of the distance, shown in Fig.2, consists of two
branches.
\begin{figure}[!htb]
\centering \centering
\includegraphics[scale=0.46, clip=true]{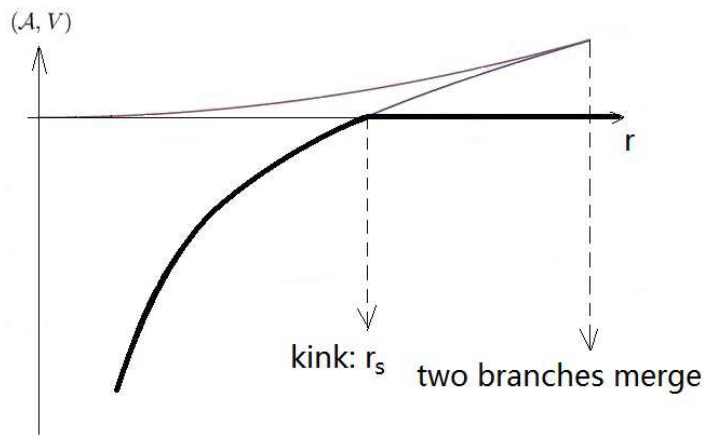}
\hspace{0in} \caption{The candidate potential (thin line) and the
heavy quark potential (thick line) of the super Yang-Mills. Both
coincide below the r-axis.}
\end{figure}
As $z_c$ starts from the AdS boundary, ${\cal A}$ starts with an
attractive Coulomb like form
\begin{equation}
{\cal A}\simeq-\frac{4\pi^2\sqrt{\lambda}}{\Gamma^4(1/4)r}
\label{Coulomb}
\end{equation}
for $rT<<1$ along the lower branch. Then ${\cal A}$ becomes
positive at $r=0.745/(\pi T)\equiv r_s$ and reaches the end of the
lower branch which corresponds to the maximum of $r$ as a function
of $z_c$. Beyond this value of $z_c$, the potential follows the
upper (repulsive) branch and decreases to zero as $z_c\to z_h$
($r\to 0$). According to (\ref{kink}), the potential $V(r)$ is
given by ${\cal A}$ for $r<r_s$ and vanishes beyond $r_s$.
Numerically, this screening potential can be well approximated by
a truncated Coulomb potential
\begin{eqnarray}
V(r)=\left\{\begin{array}{ll}
\kappa\left(\frac{1}{r}-\frac{1}{a}\right)\hbox{for $r\le a$}\\
0\ \ \hbox{for $r>a$}\end{array}\right. \label{trcoulomb}
\end{eqnarray}
with $\kappa=-\frac{4\pi^2\sqrt{\lambda}}{\Gamma^4(1/4)}$ and
$a=\frac{4\pi^2}{\Gamma^4(1/4)T}\simeq\frac{0.736}{\pi
T}$.\cite{Wu}.

\section{The Heavy-Quark Potential with a General AdS/QCD Metric at a Nonzero Temperature}

Unlike the $N=4$ super Yang-Mills, the real QCD is characterized
by an intrinsic energy scale, $\Lambda_\text{QCD}$, and the metric
of its gravity dual, if exists, should carry a length scale other
than the horizon radius. The most general metric of AdS/QCD take
the form
\begin{equation}
ds^2=\frac{w(z)}{z^2}[f(z)d\tau^2+\frac{1}{f(z)}dz^2+d\vec{x}^2]
\label{adsqcd}
\end{equation}
different proposals for the expressions of the warp factor $w(z)$
and the function $f(z)$ have been explored in the literature. They
may be simply specified to warrant an analytic
treatment\cite{Karch, Andreev4204, Andreev1} or may be dictated by
the solution of Einstein equation coupled to a dilaton field
\cite{Shock}. A number of general conditions should be satisfied
by $w(z)$ and $f(z)$: 1) The conformal invariance of QCD in UV
limit requires that $w(0)=f(0)=1$; 2) The existence of a horizon
with a nonzero Hawking temperature requires that $f(z)>0$ for
$0<z<z_h$, $f(z)=k(z_h-z)$ with $k>0$ as $z\to z_h^-$ and there is
no curvature singularity for $0\le z\le z_h$. We further assume
that both $w(z)$ and $f(z)$ are infinitely differentiable and
nonvanishing outside the horizon, $0<z<z_h$. Minimizing the
Nambu-Goto action of the string world sheet embedded in the
background metric (\ref{adsqcd}), we find the parametric form of
the function ${\cal A}(r)$
\begin{equation}
\left\{\begin{array}{ll}
r=2\sqrt{F_c}\int_0^{z_c}\frac{dz}{\sqrt{f(F-F_c)}}\\
{\cal
A}=\frac{\sqrt{\lambda}}{\pi}\bigg[\int_0^{z_c}dz\sqrt{\frac{F}{f}}\left(\frac{1}{\sqrt{1-\frac{F_c}{F}}}-1\right)
-\int_{z_c}^{z_h}dz\frac{h}{z^2}\bigg]\end{array}\right.
\label{V_eq}
\end{equation}
where
\begin{equation}
F\equiv\frac{w^2(z)}{z^4}f(z) \label{eq:F}
\end{equation}
and $F_c=F(z_c)$. The reality of the potential requires that $z_c$
stays within the domain adjacent to the boundary where $F(z)$ is
non-increasing. As $z_c\to 0$, we have $r\to 0$ and end up with
the Coulomb like potential (\ref{Coulomb}).

As a necessary condition for a smooth screening behavior, there
must exists a $z_c$ where $r\to\infty$ while $\cal A$ stays
finite. As we shall see that this is not the case. The integral
(\ref{V_eq}) diverges for $z_c=z_h$ but the factor $F_c$ removes
the divergence of the limit $z_c\to z_h$. Consequently, the candidate potential ${\cal A}$ will be restricted
within a finite $r$ like that of the super Yang-Mills, if the derivative of $F(z)$ is nonvanishing for $0<z\le z_h$.
Alternatively, the integral may also
diverge when $z_c<z_h$ but close to the point where the derivative
of $F(z)$ vanishes. But in this case, both $r$ of (\ref{V_eq}) and
${\cal A}$ of (\ref{V_eq}) share the same divergence and a linear
candidate potential at large distance emerges. A special example of the
latter case, $f(z)=1-\frac{z^4}{z_h^4}$ and $w(z)=e^{\frac{1}{2}cz^2}$ was proposed
in \cite{Andreev4204, Andreev1}to generate the Cornell potential
at low temperature \footnote{At $T=0$, the combination ${\cal A}$
diverges because of the limit $z_h\to\infty$ of the integration of
${\cal A}_1$. This reflects the nonexistence of an isolated heavy
quark because of the color confinement. At a nonzero $T$, however,
${\cal A}$ becomes finite and there is always a $r_s$ where ${\cal
A}$ switch sign. In another word, the linearly confining potential
is flattened out beyond $r_s$. There is no a clear cut distinction
between a confined phase and a plasma phase in this regard. The
deconfinement transition corresponds to a crossover from
$r_s\sim\frac{T^3}{c^2}\exp\left(\frac{c}{2\pi^2T^2}\right)>>\frac{1}{T}$
at low temperature, $T<<\sqrt{c}$, to $r_s\sim\frac{1}{T}$ at high
temperature, $T>>\sqrt{c}$.}.

To elaborate the above statements, let us examine the behavior
near the horizon, where $F(z)$ vanishes according the power law
$F(z)\simeq K(z_h-z)^n$ with $K>0$ and $n>0$ (monotonically
decreasing). It follows from (\ref{eq:F}) that the warp factor
$w(z)\sim (z_h-z)^{\frac{n-1}{2}}$. For the super Yang-Mills,
$n=1$ and $K=\frac{4}{z_h^5}$. It is straightforward to calculate
the Riemann tensor of the metric (\ref{adsqcd}). Of particular
interest is the component
\begin{equation}
R_{ijkl}=-e^{2\phi}f\phi^{\prime
2}(\delta_{ik}\delta_{jl}-\delta_{il}\delta_{jk})
\end{equation}
where $e^{2\phi}\equiv\frac{w(z)}{z^2}$ and the prime denotes the
derivative with respect to $z$. It will contributes a term
\begin{equation}
12e^{-4\phi}f^2\phi^{\prime 4} \label{singularity}
\end{equation}
to the invariant $R^{\mu\nu\rho\lambda}R_{\mu\nu\rho\lambda}$. It
is straightforward to verify that this term diverges at $z=z_h$
for all $n>0$ except that $n=1$. This divergence will lead
$R^{\mu\nu\rho\lambda}R_{\mu\nu\rho\lambda}$ to diverge because
the contribution from all components are positive. Therefore we
are left with only the case $n=1$ to consider, which gives
qualitatively the same behavior of the potential as the super
Yang-Mills in the limit $z_c\to z_h$.

To isolate out the leading behavior as $z_c\to z_h$, We introduce
$\delta$ such that $z_c-z_h<<\delta<<z_c$ and divide the
integration domain $(0,z_c)$ into $(0,z_c-\delta)$ and
$(z_c-\delta,z_c)$. In the latter domain, we may make the
approximation $f(z)\simeq k(z_h-z)$ and $F(z)=K(z_h-z)$ with $K$
another constant. Consequently
\begin{equation}
\int_{z_c-\delta}^{z_c}\frac{dz}{\sqrt{f(F-F_c)}}\simeq\frac{1}{\sqrt{kK}}\int_{f_c}^{k\delta}\frac{df}{\sqrt{f(f-f_c)}}
\simeq \frac{1}{\sqrt{kK}}\ln\frac{k\delta}{f_c}.
\end{equation}
In the former domain, we may set $F_c=0$ in the integrand and end
up with
\begin{equation}
\int_0^{z_c-\delta}\frac{dz}{\sqrt{f(F-F_c)}}\simeq\int_0^{z_h-\delta}\frac{dz}{\sqrt{fF}},
\end{equation}
which is independent of $f_c$. It follows that
\begin{equation}
r\simeq
2\frac{\sqrt{f_c}}{k}\left(\ln\frac{1}{f_c}+\text{const.}\right)
\end{equation}
as $z_c\to z_h$. Applying the same procedure to the first integral
in (\ref{V_eq}), we obtain the leading behavior
\begin{equation}
{\cal A}\simeq\frac{\sqrt{\lambda}w_c}{\pi
kz_c^2}f_c\left(\ln\frac{1}{f_c}+\text{const.}\right)
\end{equation}
which returns to zero from the positive side. Because of the
continuity of $r$ and ${\cal A}$ as functions of $z_c$, there
exists a special value of $z_c$, $z_s\in (0, z_h)$, where the
potential changes from attractive for $z_c<z_s$ to repulsive for
$z_c>z_s$. If the distance $r$ as a function of $z_c$ consists of
$m$ local maxima outside the horizon, the candidate potential
function ${\cal A}(r)$ will consists of $m+1$ branches and we have
$m\ge 1$ for the super Yang-Mills. The potential with three
branches is shown in Fig.3. To reproduce the kink-like screening
of Fig.2, the potential ${\cal A}$ as a function of $z_c$ has to
be positive at the first local maximum of $r(z_c)$. The heavy
quark potential will follow ${\cal A}(r)$ for $z<z_s$ and becomes
flat afterwards.
\begin{figure}[!htb]
\centering \centering
\includegraphics[scale=0.42, clip=true]{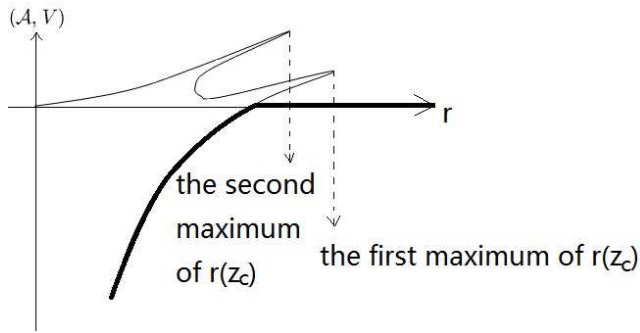}
\hspace{0in} \caption{An example of the candidate potential ${\cal
A}$ with three branches (thin line) together with the
corresponding kink-like heavy quark potential $V$ (thick line) of
AdS/QCD.}
\end{figure}

Next, we explore the case where the derivative of $F(z)$ vanishes
somewhere outside the horizon, i.e. $F'(z_0)=0$ for $0<z_0<z_h$.
As $z\to z_0$, we may approximate
\begin{equation}
F(z)\simeq a+b(z_0-z)^n
\end{equation}
with positive $a$ and $b$, and an integer $n\ge 2$. For an even
$n$, $z_c$ has to stay on the side $z_c<z_0$ in order for the
square root to be real. As $z_c\to z_0$, both the integral in
(\ref{V_eq}) and the first integral for (\ref{V_eq}) diverges.
Dividing the integration domain $(0,z_c)$ into $(0,z_c-\delta)$
and $(z_c-\delta,z_c)$ with
$\epsilon\equiv|z_0-z_c|<<\delta<<z_0$, we can easily extract the
leading behaviors for $z_c\to z_0-0^+$,
\begin{equation}
\left\{\begin{array}{ll}
r\simeq\frac{2w_0}{\sqrt{b}z_0^2}\left(\ln\frac{z_0}{\epsilon}+\text{const.}\right)\\
{\cal
A}\simeq\sqrt{\frac{\lambda}{bf_0}}\frac{a}{\pi}\left(\ln\frac{z_0}{\epsilon}+\text{const.}\right)\end{array}\right.
\end{equation}
for $n=2$ and
\begin{equation}
\left\{\begin{array}{ll}
r\simeq \frac{2}{n}B\left(\frac{1}{2}-\frac{1}{n},\frac{1}{2}\right)\sqrt{\frac{a}{bf_0}}\epsilon^{1-\frac{n}{2}}\\
{\cal A}\simeq
\frac{1}{n\pi}B\left(\frac{1}{2}-\frac{1}{n},\frac{1}{2}\right)\sqrt{\frac{\lambda
a^2}{bf_0}}\epsilon^{1-\frac{n}{2}}\end{array}\right.
\end{equation}
for $n>2$. Similar leading behavior as $z_c\to z_0+0^+$ read
\begin{equation}
\left\{\begin{array}{ll} r\simeq
\frac{2}{n}\bigg[B\left(\frac{1}{n},\frac{1}{2}\right)+B\left(\frac{1}{n},\frac{1}{2}-\frac{1}{n}\right)\bigg]
\sqrt{a}{bf_0}\epsilon^{1-\frac{n}{2}}\\
{\cal A}\simeq
\frac{1}{n\pi}\bigg[B\left(\frac{1}{n},\frac{1}{2}\right)+B\left(\frac{1}{n},\frac{1}{2}-\frac{1}{n}\right)\bigg]
\sqrt{\lambda a^2}{bf_0}\epsilon^{1-\frac{n}{2}}\end{array}\right.
\end{equation}
for an odd $n\ge 3$. In all cases above, the function ${\cal
A}(r)$ contains a branch of Cornell potential. The examples of
$F(z)$ and the corresponding ${\cal A}(r)$ and $V(r)$ for an even
$n$ and for an odd $n$ are depicted in Figs. 4-5.
\begin{figure}[!htb]
\centering \centering
\includegraphics[scale=0.8, clip=true]{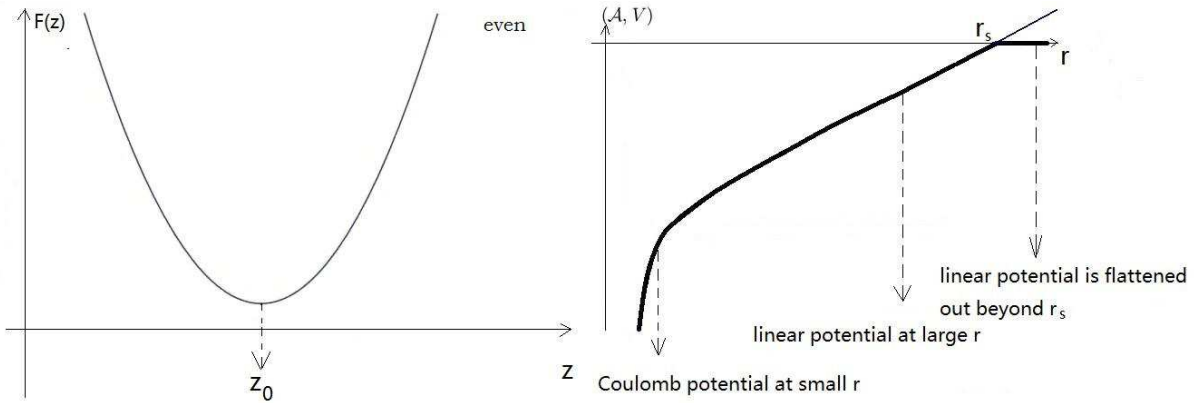}
\hspace{0in} \caption{The left panel shows the example of F(z)
whose expansion around $z_0$ is $F(z)\simeq a+b(z_0-z)^n$ with an
even $n$. We marked $z_0$ with dashed line. The right panel is the
corresponding candidate potential ${\cal A}$ as a function of r
which is a Coulomb potential for small r combining with a linear
potential at large r limit. The thick line represents the heavy
quark potential $V$.}
\end{figure}
\begin{figure}[!htb]
\centering \centering
\includegraphics[scale=0.82, clip=true]{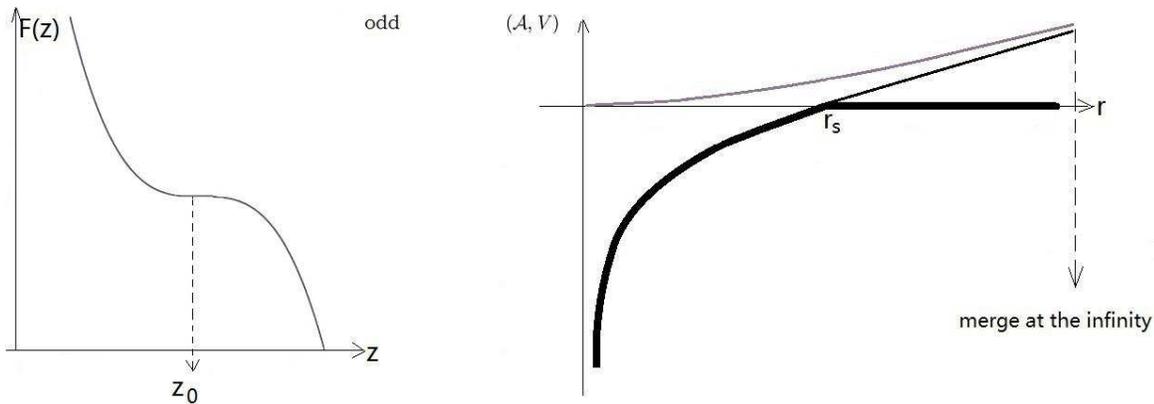}
\hspace{0in} \caption{The left panel shows the example of F(z)
whose expansion around $z_0$ is $F(z)\simeq a+b(z_0-z)^n$ with an
odd $n$. The right panel shows the two branches of the candidate
potential ${\cal A}$ as a function of r. Both branches
rise linearly with $r$ and will merge at $r\rightarrow\infty$. The
thick line represents the heavy quark potential $V$.}
\end{figure}

\section{Discussions}

In previous sections, we have explored all possible forms of the
holographic heavy-quark potential at a nonzero temperature with a
general metric that is asymptotically AdS and carries a black. The
candidate potential function ${\cal A}(r)$ is either supported
within a finite range of the distance $r$ (Figs.2-3) or takes the
linear form for large $r$ when $r$ is allowed to go to infinity
(Figs.4-5). The heavy quark potential $V(r)$ coincide with ${\cal
A}$ when the free energy of the an interacting quark-antiquark
pair (represented by the world sheet of Fig.1a) is lower than that
of a non-interacting pair (represented by the world sheet of
Fig.1b) and is flattened out otherwise. A kink-like screening
behavior with a discontinuity in the derivative $dV/dr$ is
developed then.

In this section we would like first to comment on some proposals
in the literature to smooth out the kink-like the heavy quark
potential in a thermal bath within the framework of the super
Yang-Mills. The authors of \cite{Kovchegov} suggested that the
parameter $z_c$ becomes complex beyond the value when $r$ is
maximized. So the potential develops an imaginary part and decays
with power law. This, however, cannot be the case at thermal
equilibrium using the definition(\ref{def})   since the thermal expectation value $\text{Tr}<{\cal
P}(\vec r){\cal P}^\dagger(0)>$ is strictly real as is evident
from the following reasoning:
\begin{equation}
\text{Tr}<{\cal P}(\vec r){\cal P}^\dagger(0)>^*=\text{Tr}<{\cal
P}(0){\cal P}^\dagger(\vec r)> =\text{Tr}<{\cal P}(-\vec r){\cal
P}^\dagger(0)>=\text{Tr}<{\cal P}(\vec r){\cal P}^\dagger(0)>
\end{equation}
where the second equality follows from the translation invariance
and the last equality from the rotation invariance (the thermal
expectation value should be a function of $|\vec r|$).  One can obtain a complex potential using other
definition  or by analytic continuation\cite{Hatsuda-lattice, Hatsuda}.
 The authors of \cite{Yaffe} suggested an alternative string world sheet by
joining the two parallel world sheets of noninteracting quarks
with a thin tube to represent the exchange of the lightest
supergravity mode. While physically plausible, the area of such a
configuration is always larger than that without the tube because
the background metric (\ref{Sch_AdS}) is {\it independent} of the
transverse coordinates $\vec x$.

Next, we shall question the legitimacy of a kink-like screening
potential from field theoretic point of view. Because of the
$O(4)$ symmetry of the Lagrangian density, the path integral of a
thermal relativistic field theory in $R^3$ at temperature $T$, is
mathematically equivalent to the field theory under the same
Lagrangian density but formulated in $R^2\times S^1$ at zero
temperature with a Euclidean time \cite{Yaffe,Liu}, provide one of
the spatial dimensions of the former, say $x^3$ is interpreted as
the Euclidean time of the latter and the Euclidean time of the
former is regarded as the compactified spatial dimension. The
latter will be referred to as the adjoint field theory and is as
well defined as the original thermal field theory. Consequently, a
static Green's function, such as the heavy-quark potential of the
thermal field theory, becomes a time-dependent Green's function of
the adjoint field theory and the analyticity on the complex energy
plane of its Fourier transformation should meet the requirements
imposed by the unitarity and causality. Furthermore, the Fourier
transformation should vanish at infinity of the energy plane in
order for the Fourier integral of the retarded (advanced) Green's
function exists. Let us examine if this is the case for a
potential $V(r)$ that is continuous and vanishes for
$r>r_s$\footnote{Here we assume that $\frac{dV}{dr}\ne 0$ at
$r=r_s$ but this is not necessary for our conclusion. If the
derivatives of $V$ up to $n$-th order vanishes at $r_s$, we may
conduct the integration by part $n$ times more in the process of
(\ref{exponential}) below and the exponential factor $e^{\omega
r_s}$ remains.}. The Fourier transformation
\begin{equation}
{\cal V}(q)=\int d^3\vec{r} e^{-i\vec q\cdot\vec r}V(r) =
\frac{4\pi}{q}\int_0^{r_s} drrV(r)\sin qr \equiv {\cal
G}(iq_3,\vec q_{\perp})
\end{equation}
where $q=\sqrt{q_3^2+\vec q_\perp^2}=\sqrt{-\omega^2+\vec
q_\perp^2}$ with $\omega=iq_3$. For a given $\vec q_\perp$ and a
sufficiently large real energy, we need to continuate the integral
to an imaginary $q$ and  this can be done at the integrand level because of the finite integration
limits. We have
\begin{eqnarray}
{\cal G}(\omega,\vec q_{\perp}) &=& \frac{4\pi}{\sqrt{\omega^2-\vec q_{\perp}^2}}\int_0^{r_s} dr r V(r)\sinh\sqrt{\omega^2-\vec q_\perp^2}r\\
&=&-\frac{4\pi}{\omega^2-\vec q_\perp^2}\lim_{r\to
0}rV(r)-\frac{4\pi}{\omega^2-\vec q_\perp^2}
\int_0^{r_s} dr\frac{d}{dr}(r V)\cosh\sqrt{\omega^2-\vec q_{\perp}^2}r\\
&\simeq& -\frac{2\pi}{\omega^3}e^{\omega r_s}r_sV^\prime(r_s)
\label{exponential}
\end{eqnarray}
as $\omega\to\infty$, where an integration by part is conducted
from the first line to the second line. A concrete example is the
truncated Coulomb potential (\ref{trcoulomb}) and the
corresponding
\begin{equation}
{\cal G}(\omega,\vec q_\perp)=\frac{4\pi\kappa}{\omega^2-\vec
q_\perp^2}[1-i_0(\sqrt{\omega^2-\vec q_\perp^2}r_s)]
\end{equation}
with $i_0(z)=\sinh z/z$. While there are no complex singularities
in $\omega$-plane as is required by the unitarity, the
exponentially growing behavior of ${\cal G}(\omega,\vec q_\perp)$
at large $\omega$ prevents us from defining the retarded and
advance Green's function of the adjoint field theory via Fourier
integrals
\begin{equation}
G_{R(A)}(t,\vec
q_\perp)\equiv\int_{-\infty}^\infty\frac{d\omega}{2\pi}e^{-i\omega
t}{\cal G}(\omega\pm i0^+,\vec q_\perp)
\end{equation}
with the upper(lower) sign for $R(A)$. We regard this problem a
potential gap between the holographic approach and the field
theoretic principle. It requires further investigations, perhaps
some resummation of the finite $N_c$ or and/or finite $\lambda$
corrections to the leading form (\ref{leading}) to smear the kink of in
the screening potental.

\begin{acknowledgments}
The research of Defu Hou and Hai-cang Ren is supported in part by
NSFC under grant Nos. 11375070 ,  11221504, 11135011. Yan Wu is
partly supported by excellent doctorial dissertation cultivation
grant from Central China Normal University under No. 2013YBYB45.
\end{acknowledgments}

\end{document}